\documentstyle[12pt]{article}

\def\etal{\hbox{et al.}{}} 	

\def\cm{\hbox{c.m.}{}}

\def\ltap{\;\raisebox{-.5ex}{\rlap{$\sim$}} \raisebox{.5ex}{$<$}\;}

\def\r#1{\ignorespaces $^{#1}$}

\def\beq{\begin{equation}}
\def\eeq{\end{equation}}

\def\beqn{\begin{eqnarray}}
\def\eeqn{\end{eqnarray}}

\relax

\catcode`\@=11

\@addtoreset{equation}{section}
\def\theequation{\thesection\arabic{equation}}

\def\@normalsize{\@setsize\normalsize{15pt}\xiipt\@xiipt
\abovedisplayskip 14pt plus3pt minus3pt%
\belowdisplayskip \abovedisplayskip
\abovedisplayshortskip \z@ plus3pt%
\belowdisplayshortskip 7pt plus3.5pt minus0pt}

\def\small{\@setsize\small{13.6pt}\xipt\@xipt
\abovedisplayskip 13pt plus3pt minus3pt%
\belowdisplayskip \abovedisplayskip
\abovedisplayshortskip \z@ plus3pt%
\belowdisplayshortskip 7pt plus3.5pt minus0pt
\def\@listi{\parsep 4.5pt plus 2pt minus 1pt
     \itemsep \parsep
     \topsep 9pt plus 3pt minus 3pt}}

\relax

\catcode`@=12

\catcode`\@=11

\def\section{\@startsection{section}{1}{\z@}{3.5ex plus 1ex minus .2ex}
{2.3ex plus .2ex}{\large\bf}}

\def\thesection{\arabic{section}.}

\def\appendix{\setcounter{section}{0}
 \def\thesection{APPendIX \Alph{section}:}
 \def\theequation{\Alph{section}.\arabic{equation}}}

\relax
\def\pl#1#2#3{Phys. Lett. {\bf #1} (19#2) #3}
\def\zp#1#2#3{Z. Phys. {\bf #1} (19#2) #3}
\def\prl#1#2#3{Phys. Rev. Lett. {\bf #1} (19#2) #3}

\def\prev#1#2#3{Phys. Rev. {\bf #1} (19#2) #3}
\def\np#1#2#3{Nucl. Phys. {\bf #1} (19#2) #3}

\relax

\begin{document}


\newcommand{\ttb}{ttb}
\newcommand{\sss}{\scriptscriptstyle}
\newcommand{\rar}{\rightarrow}
\newcommand{\ve}{$\nu_{\sss e}$}
\newcommand{\epem}{$e^{\sss +}e^{\sss -}$}
\newcommand{\mt}{m_{\sss t}}
\newcommand{\mh}{m_{\sss H}}
\newcommand{\mz}{M_{\sss Z}}
\newcommand{\mw}{M_{\sss W}}
\newcommand{\mb}{m_{\sss b}}
\newcommand{\rs}{\sqrt{s}}
\newcommand{\mwb}{M_{\sss Wb}}
\newcommand{\cost}{\cos \theta}
\newcommand{\gtop}{\Gamma_{t}}
\newcommand{\gw}{\Gamma_{\sss W}}
\newcommand{\gz}{\Gamma_{\sss Z}}
\newcommand{\gele}{\Gamma_{e}}
\newcommand{\gfer}{\Gamma_{f}}
\newcommand{\glep}{\Gamma_{\ell}}
\newcommand{\ghad}{\Gamma_{\sss had}}
\newcommand{\gbot}{\Gamma_{b}}
\newcommand{\ginv}{\Gamma_{\sss INV}}
\newcommand{\gaa}{g_{\sss Af}}
\newcommand{\gvv}{g_{\sss Vf}}
\newcommand{\gab}{g_{\sss Ab}}
\newcommand{\gvb}{g_{\sss Vb}}
\newcommand{\gac}{g_{\sss Ac}}
\newcommand{\gvc}{g_{\sss Vc}}
\newcommand{\gal}{g_{\sss A\ell}}
\newcommand{\gvl}{g_{\sss V\ell}}
\newcommand{\gga}{g_{\sss A}}
\newcommand{\ggv}{g_{\sss V}}

\newcommand{\rl}{R_{\ell}}
\newcommand{\rb}{R_{b}}
\newcommand{\rd}{R_{d}}
\newcommand{\re}{R_{e}}
\newcommand{\rmu}{R_{\mu}}
\newcommand{\rtau}{R_{\tau}}

\newcommand{\alr}{A_{\sss LR}}
\newcommand{\afb}{A_{\sss FB}}
\newcommand{\afbf}{A^f_{\sss FB}}
\newcommand{\afbl}{A^{\ell}_{\sss FB}}
\newcommand{\afbe}{A^{e}_{\sss FB}}
\newcommand{\afbmu}{A^{\mu}_{\sss FB}}
\newcommand{\afbtau}{A^{\tau}_{\sss FB}}
\newcommand{\afbb}{A^{b}_{\sss FB}}
\newcommand{\afbc}{A^{c}_{\sss FB}}

\newcommand{\ptau}{P_{\tau}}
\newcommand{\ptaufb}{P_{\tau}^{\sss FB}}
\newcommand{\qfb}{Q_{\sss FB}}

\newcommand{\sigff}{\sigma_{\sss \bar{f}f}}
\newcommand{\sigffcap}{\hat{\sigma}_{\sss \bar{f}f}}
\newcommand{\sigllcap}{\hat{\sigma}_{\sss \bar{\ell}\ell}}
\newcommand{\sigoh}{\sigma^{\sss 0}_{\sss had}}
\newcommand{\sigol}{\sigma^{\sss 0}_{\sss \ell\ell}}
\newcommand{\sigob}{\sigma^{\sss 0}_{\sss \bar{b}b}}
\newcommand{\sigoff}{\sigma^{\sss 0}_{\sss \bar{f}f}}

\newcommand{\abar}{\overline{\alpha}}
\newcommand{\alfs}{\alpha_{\sss s}(\mz)}
\newcommand{\alfspi}{\frac{\alpha_{\sss s}(\mz)}{\pi}}

\newcommand{\st}{\sin\theta_{\sss w}}
\newcommand{\ct}{\cos\theta_{\sss w}}
\newcommand{\stq}{\sin^{2}\theta_{\sss w}}
\newcommand{\ctq}{\cos^{2}\theta_{\sss w}}
\newcommand{\steff}{\sin^{2}\theta_{\sss w}^{\sss eff}}
\newcommand{\roff}{\rho^{\sss eff}}
\newcommand{\Wpm}{$W^{\sss \pm}$}
\newcommand{\Z}{$Z^{\sss 0}$}
\newcommand{\epuno}{$\varepsilon_1$}
\newcommand{\epdue}{$\varepsilon_2$}
\newcommand{\eptre}{$\varepsilon_3$}
\newcommand{\mepb}{$\varepsilon_b$}

\setcounter{topnumber}{10}
\setcounter{bottomnumber}{10}
\renewcommand\topfraction{1}
\renewcommand\textfraction{0}
\renewcommand\bottomfraction{1}

\relax


\begin{titlepage}
\pagestyle{empty}
\begin{flushleft}
{\bf Preprint n.984} \hfill Dipartimento di Fisica \ \ \ \ \ \ \ \ \ \\
December 14, 1993 \hfill Universit\`a di Roma ``La Sapienza'' \\
\hfill I.N.F.N. - Sezione di Roma \ \ \ \ \ \  \\
\end{flushleft}
\vspace*{\fill}
\begin{center}
{\Large \bf Precision Tests of the Standard Model at LEP}

\vspace*{\fill}

\begin{tabular}[t]{c}
{\large Barbara Mele\r{1}} \\
\\
{ \it I.N.F.N., Sezione di Roma, Italy \ and } \\
{ \it Dipartimento di Fisica, Universit\`a ``La Sapienza'',} \\
{ \it P.le Aldo Moro 2, I-00185 Rome, Italy} \\
\end{tabular}
\end{center}
\vspace*{\fill}
\begin{abstract}
{\small
\noindent
Recent LEP results on electroweak precision measurements are reviewed.
Line-shape and asymmetries analysis on the \Z\ peak is described.
Then, the consistency of the Standard Model predictions with experimental
data and consequent limits on the top mass are discussed. Finally,
the possibility of extracting information and constrains on new
theoretical models from present data is examined. }
\end{abstract}
\vspace*{\fill}
{\small
{\it Invited talk at the XIV Encontro Nacional de Fisica de Campos
e Particulas, Caxambu,
Brasil,  29 Sept - 3 Oct, 1993}.
\\
\r{1}\ {\it e-mail address: MELE@ROMA1.INFN.IT}}
\end{titlepage}

\section{Introduction}

\pagestyle{myheadings}

Four years of LEP operation, from the machine starting in 1989,
have yielded an impressive amount of
new data describing physics at the \Z\ peak.
In the years '89-'92, the four LEP experiments
(ALEPH, DELPHI, L3 and OPAL) have collected
 about 5 million of \Z\ events which have already
 been analysed \cite{pre}.
These data have allowed up to now the measurements of many observables
in the Electro-Weak (EW) sector of the Standard Model to a much
higher precision than before the LEP era.
The running of the machine during two further years will yield by the end
of '94 about four times the present statistics, and will improve even
more the present achievements.

In my talk, I would like to review the beautiful LEP results
in the EW sector of the Standard Model, concentrating on precision
tests of the SU(2)$\times$U(1) theory at $\rs=\mz$. I will describe
the  main strategies to extract precision measurements from LEP
experimental data, through the \Z\ line-shape analysis and asymmetries
determination on the \Z\ peak. Then, I will discuss the experimental
consistency of the Standard Model and the present status
of precision tests of the {\it purely} EW sector of the theory.
Limits on the top mass will be discussed as well.
 Finally, I will  briefly report on the results of model
independent analysis of the precision EW data, and consequent
constraints that can be put on possible extensions of the Standard
Model starting from present LEP data. My discussion will be based
mostly on LEP results recently presented at the
Europhysics International Conference of Marseilles \cite{pre,lef} and at
the XVI International Symposium on Lepton-Photon Interactions
at Cornell University, N.Y.\cite{swa,hol}.

\section{Experimental analysis}
Due to the \Z\ peak,  one observes at LEP a bump in the
cross section at $\rs \simeq \mz$ with a peak value of
about 35 nb for {\it visible} (that is hadronic and charged-lepton)
events. For instance, the hadronic cross section
is about 750 times what one would expect by extrapolating
the low-energy purely photonic contribution up to $\rs \simeq \mz$.
The corresponding visible-event rate is about $3.5\cdot$10\r{6}
for an effective year of running of 10\r{7}s, with a luminosity of
10\r{31}cm\r{-2}s\r{-1} (that is about twice the
present typical LEP luminosity).
With such a high statistics, LEP gives a unique  opportunity to test
the {\it fine} structure of the Standard Model at the $\mz$ energy scale.
In order to fully exploit the wealth of data for precision
measurements, one has to keep under control systematic errors in
the experimental analysis. To this end, two basic issues are the
luminosity monitoring and the beam energy calibration of the machine.
 The LEP luminosity $\cal L$ is monitored through the measurements
of Bhabha \epem$\rar$\epem scattering.
Hence, the total cross section for production of the fermion
pair $\bar{f}f$ is obtained by the expression
\beq
\sigma_{\bar{f}f}=\frac{N_{\bar{f}f}}{\varepsilon_{\bar{f}f}}\cdot
\frac{1}{\int {\cal L} dt}
=\frac{N_{\bar{f}f}}{\varepsilon_{\bar{f}f}}\cdot
\frac{\varepsilon_{\sss Bhabha}}{N_{\sss Bhabha}}\cdot
\sigma^{\sss theory}_{\sss Bhabha}
\label{rate}
\eeq
where $N$ is the observed number of events and
$\varepsilon$ is a factor including acceptance and efficiency effects.
The  main limitation in the luminosity measurement comes from the
theoretical error in the prediction of the Bhabha cross section
$\sigma^{\sss theory}_{Bhabha}$.
As a consequence, $\cal L$ is presently determined
with a relative error of about $1\%$,
whose main effect reflects in limiting the accuracy of the peak
hadronic cross section $\sigma^0_{h}$ to about $0.3\%$\cite{lef}.
On the other hand, a very good beam energy calibration is now
reached through the resonant spin depolarization method, that exploits
the transverse polarization of the initial beams. \\
By the way, it is interesting to recall
that, in depolarization measurements, some energy spread has been observed
which is correlated with tidal effects, that deform the earth's surface.
As a consequence, due to the variation
of the collider circumference by a few 10\r{-8},
 a periodic change (of amplitude about 9.6 MeV)  is observed in the beam
energy measured by resonant depolarization \cite{lef}. \\
Such a good energy calibration has allowed to get
systematic errors on the \Z\ mass $\mz$ and width $\gz$ of only 6.3 MeV
and 4.5 MeV, respectively.

The precision determination of various EW observables at LEP is obtained
by elaborating two main kinds of {\it primary} measurements :
1) cross sections and 2) asymmetries.
In what follows, I will describe the main strategies to measure
these quantities.

\subsection{Cross sections}
By energy scanning around the \Z\ peak (from about $\rs=88$GeV
up to about 94 GeV),
 one measures (through eq.(\ref{rate}))
the cross section $\sigma_{\bar{f}f}(s)$, for production of
the $f$ fermion pair, versus the \cm\ energy. A Breit-Wigner
resonant shape around $\mz$ is obtained. There are four different main cases
corresponding to the three charged-lepton ($f=e,\mu,\tau$) and
to hadron ($f=\sum q_i$) production.
In order to derive from the measured cross sections a measurement of
relevant quantities (\Z\ mass, total and partial \Z\ widths and peak cross
sections), one has first to subtract the effect of initial-state
photon radiation \cite{ber}. This is a rather large effect, that causes a
reduction of about $25\%$ in the peak cross section. It can be accounted
for through a radiator function $G(z,s)$ that can be deconvoluted
from the measured $\sigma_{\bar{f}f}(s)$
\beq
\sigma_{\bar{f}f}(s)=\int dz \; G(z,s) \hat{\sigma}_{\bar{f}f}(zs)
\label{radia}
\eeq
where $z$ is the fraction of the \cm\ energy squared $s$
left after the photon radiation to the  collision,
 and $\hat{\sigma}_{\bar{f}f}(s)$ is  defined by eq.(\ref{radia}) as
a reduced cross section deconvoluted from initial radiation effects.
$G(z,s)$ can be theoretically predicted to a good accuracy
through renormalization-group
methods, that resums the effect of large terms of the order
$\frac{\alpha}{\pi}\log(\mz^2/m^2_{f_{\sss light}})$.

At this point, $\mz$ and $\gz$ are {\it defined} through the expression
\beq
\hat{\sigma}_{\sss \bar{f}f}(s)=\sigoff \cdot
\frac{s\gz^2}{(s-\mz^2)^2+(s^2\gz^2)/\mz^2} + (\gamma \, exchange
+ Inter's)
\label{breit}
\eeq
where the last term in brackets is small and takes
into account the photon exchange contribution
and its interference with the \Z\ amplitude. It is theoretically
evaluated and subtracted from $\hat{\sigma}_{\bar{f}f}(s)$, in order
to isolate the Breit-Wigner \Z\ contribution.
The peak cross section $\sigoff$ is connected to the \Z\ partial widths
for $Z\rar ee$ and $Z\rar ff$, $\gele$ and $\gfer$, through the expression
\beq
\sigoff= \frac{12\pi}{\mz^2} \frac{\gele \gfer}{\gz^2}
\label{sigoff}
\eeq
Of particular relevance for the LEP data analysis are the peak
hadronic and leptonic cross sections. According to eq.(\ref{sigoff}),
the former is directly related to the \Z\ hadronic width $\ghad$ by
\beq
\sigoh= \frac{12\pi}{\mz^2} \frac{\gele \ghad}{\gz^2}
\label{sigoh}
\eeq
The latter is replaced by the ratio of the hadronic and leptonic \Z\ widths
\beq
\rl\equiv \frac{\sigoh}{\sigol}=\frac{\ghad}{\glep}
\label{rl}
\eeq
The use of the ratio's $\rl$, with $\ell=e,\mu,\tau$,
parametrizes the \Z\ leptonic
couplings, avoiding the systematic uncertainties connected to the
measurement of cross sections.
In the Standard Model $\glep$ can be expressed in terms of
the vector and axial-vector lepton coupling constants, $\gvl$ and $\gal$, by
\beq
\glep= \frac{G_{\sss F}\mz^3}{6\pi \sqrt{2}}(\gvl^2+\gal^2)
\left(1+\frac{3}{4}\frac{\alpha}{\pi}\right)
\label{glep}
\eeq
where the last term in brackets takes into account electromagnetic
corrections.

With the above definitions, hadronic and leptonic line-shape data
are analyzed in a model-independent way
(separately by each of the four LEP experiments),
 assuming $\mz, \gz, \sigoh, \rl$ and the asymmetries $\afbl$
(defined in the  next section) as the set of most independent parameters
\cite{pre}.
A combined fit to the line-shapes and $\afb$'s is made, based on a
$\chi^2$-minimization that takes into account the full covariant error
matrix of the data, and the experimental and theoretical correlations
between different channels.
About leptonic data, two different assumptions are made.
Assuming lepton universality, one has $\re=\rmu=\rtau=\rl$ and
$\afbe=\afbmu=\afbtau=\afbl$ and fits the data to the 5-parameters
$\mz, \gz, \sigoh, \rl$ and $\afbl$. Releasing this
assumption, one makes a 9-parameter fits on
$\mz, \gz, \sigoh, \re, \rmu, \rtau, \afbe, \afbmu, \afbtau$.
In Table 1 \cite{lef}, the values obtained from a 5-parameter fit, after
combining the results from the four LEP experiments, are presented.
\begin{table}

\[ \begin{array}{|c||c|} \hline
\mz \: (GeV) & 91.187 \pm 0.007 \\ \hline
\gz \: (GeV) & 2.489 \pm 0.007 \\ \hline
\sigoh \: (nb) & 41.55 \pm 0.14 \\ \hline
\rl=\ghad/\glep & 20.77 \pm 0.05 \\ \hline
\rb=\gbot/\ghad & 0.2191 \pm 0.0027 \\ \hline
\end{array} \]  \label{fit5}

\caption{Primary measurements (apart from asymmetries).}

\end{table}
One can remark the impressive precision obtained
on $\mz$. Its relative error is about $8\cdot10$\r{-5} and is going
to improve even further after the '93 data. The improvement in
energy calibration will lower the present systematic error of $\pm0.006$
down to $\pm0.0025$. For $\gz$ the energy-scale error will go
from $\pm0.0045$ down to $\pm0.002$, allowing an accuracy better than
the present $0.3\%$. The main systematics for $\sigoh$ comes instead
from the luminosity monitoring and, therefore, from theoretical
uncertainties on $\sigma^{\sss theory}_{\sss Bhabha}$.

{}From the measurement of the {\it primary} quantities $\mz, \gz, \sigoh$
and $\rl$, one can
straightforwardly obtain a measure of some {\it derived} quantities,
that are the \Z\ leptonic and hadronic widths, $\glep$ and $\ghad$,
and the \Z\ {\it invisible} width $\ginv$. The last is directly connected
with the number of light neutrino species.
{}From eqs.(\ref{sigoh}) and (\ref{rl}) (assuming from lepton universality
$\gele=\glep$), starting from the measured ÿ$\mz, \gz, \sigoh$
and $\rl$, the combined four LEP experiment results are \cite{lef}
\beq
\glep = (83.79\pm0.28) MeV ; \;\;\;\;\;\;\;\; \ghad = (1740\pm6) MeV
\label{gg}
\eeq
The invisible \Z\ width is defined as the difference between the total
 \Z\ width and the sum of all the \Z\ {\it visible} decay widths
\beq
\ginv=\gz-\ghad-3\glep
\label{ginv}
\eeq
Assuming that only Standard Model $\nu$'s contributes to $\ginv$,
the value of $\ginv$ is proportional to $N_{\nu}$, the number of neutrino's
lighter than $\mz/2$. Actually, the Standard Model gives a cleaner
prediction for the ratio $\ginv/\glep$, since EW corrections in each
single width, that are dependent on the unknown $\mt$ and $\mh$,
largely  cancel in the ratio. From the Standard Model, one has \cite{lef}
\beq
\frac{\ginv}{\glep}=(1.994\pm0.003)N_{\nu}
\label{nnu1}
\eeq
On the other hand, the above ratio can be experimentally determined
from primary measurements through
\beq
\frac{\ginv}{\glep}=\sqrt{\frac{12\pi \rl}{\mz^2\sigoh}}-\rl-3
\label{nnu2}
\eeq
where eqs.(\ref{sigoh}) and (\ref{rl}) have been used in  eq.(\ref{ginv}).
Assuming a central value for the coefficient in eq.(\ref{nnu1}), one gets,
 combining the four-experiment results
\beq
N_{\nu}=2.980\pm0.027
\eeq
LEP confirms with a remarkable (and unprecedented) accuracy the minimal
Standard Model prediction of 3 families of light $\nu$'s.

A {\it derived} quantity of different nature from EW data is the measurement
of the QCD strong
coupling constant at the $\mz$ scale, $\alfs$. Since QCD corrections
affect considerably $\ghad$, the values of both $\rl$ and $\gz$
are  sensitive to $\alfs$. $\rl$ is the most sensitive variable.
 Its dependence on $\alfs$ has been calculated up to
the third order in the perturbative expansion
\beq
\rl=\rl^{\sss 0} \left\{1+1.05\left(\alfspi\right)+
(0.9\pm0.1)\left(\alfspi\right)^2-13\left(\alfspi\right)^3\right\}
\eeq
{}From the $\rl$ measurement alone, one gets a value \cite{lef}
\beq
\alfs=0.123\pm0.008
\eeq
while a combined fit of all EW observables to $\alfs$ gives
(cf. section 3)
\beq
\alfs=0.120\pm0.007
\label{alfs}
\eeq
Comparing this determination with its  accuracy to the results
of hadronic-event-shape analysis ($\alfs=0.123\pm0.006$),
one finds perfect agreement and gets a nice consistency check
of the SU(3)$\times$SU(2)$\times$U(1) theory.
The accuracy in eq.(\ref{alfs}) is essentially limited by our ignorance
of $\mt$ and $\mh$, which enter the EW corrections that must be subtracted
before fixing QCD corrections.

Recently, the implementation of a micro-vertex detector for
$b$-tagging at LEP, has allowed  a rather accurate measurement of
the total cross section for $Z\rar\bar{b}b$ at $\rs=\mz$, and in particular
of the ratio of the $b$ over the hadronic \Z\ width
\beq
\rb\equiv \frac{\sigob}{\sigoh}=\frac{\gbot}{\ghad}
\label{rb}
\eeq
This ratio has the advantage of being rather sensitive to top-quark
vertex corrections ($t$ enters the 1-loop $Z\rar\bar{b}b$ diagrams without
Cabibbo-Kobayashi-Maskawa suppression) and, at the same time,
less sensitive than the individual $\gbot$ and $\ghad$  to
the exact value of $\alfs$ that heavily affects all the  hadronic widths
through QCD radiative corrections. In the Standard Model, one has
\cite{ver}
\beq
\rb\simeq\rd\left\{1-\frac{20}{13}\frac{\alpha}{\pi}\left(
\frac{\mt^2}{\mz^2}+\frac{13}{6}\log(\frac{\mt^2}{\mz^2})\right)\right\}
\label{rb2}
\eeq
where $\rd$ is the analogous of $\rb$ for $d$ quarks (for which top-loop
effects are negligible). If $\mt=150$GeV, eq.(\ref{rb2}) gives a $2\%$
effect. Therefore, if $\rb$ is measured with an accuracy better than
about $1\%$, one can constrain $\mt$. Contrary to other EW observables
at LEP, this effect is rather clean, since it is almost independent on
$\mh$ and other \Z\ propagator effects.
The measured value of $\rb$ (reported in Table 1)
puts an upper limit of 221 GeV on $\mt$
at 95$\%$ of confidence level (using also the present Tevatron lower
limit on $\mt$ of 113 GeV)\cite{lef}. This strategy for limiting the top
mass is thus becoming competitive with the results of global fits of LEP
data to the Standard Model (cf. section 3).

\subsection{Asymmetries}
There are different kinds of asymmetries measured at LEP\cite{akv}:
forward-backward asymmetries for charged fermions (leptons and heavy
quarks), $\afbf$, $\tau$-polarization asymmetries, $\ptau$,
$\tau$-polarization forward-backward asymmetries, $\ptaufb$,
charge forward-backward asymmetries, $\qfb$.
 The main goal of determining these quantities is an accurate measurement
of the ratio of vector
and axial \Z\ couplings to fermions and, as a consequence, of the value
of $\steff$.
This quantity is a basic one in order to check the radiative-correction
pattern of the Standard Model.
In practice, all these asymmetries can be expressed
as functions of the quantities $A_f$'s, that express
the unbalance in the  left- and right-handed
fermion couplings to the \Z\
\beq
A_f\equiv\frac{g^2_{\sss L}{\sss(f)}-g^2_{\sss R}{\sss(f)}}
{g^2_{\sss L}{\sss(f)}+g^2_{\sss R}{\sss(f)}}
=\frac{2\gvv\gaa}{\gvv^2+\gaa^2}={\cal F}\left(\frac{\gvv}{\gaa}\right)
\label{af}
\eeq
where $f=e,\mu,\tau,c,b$ for the experimentally interesting cases,
and $\ggv$ and $\gga$  enters the $Zf\bar{f}$ coupling
$\gamma^{\mu}(\gvv-\gaa\gamma_5)$. In the following, whenever $\gvv$ and
$\gaa$ are reported without a suffix $f$, they refer to the leptonic
couplings (assuming lepton universality for $e,\mu$ and $\tau$).
In fact, leptonic asymmetries are the easiest to determine experimentally
due to the cleaner reconstruction of leptonic final states.

In the Standard Model the vector and axial-vector couplings for leptons
at the \Z\ peak can be parametrized in the following way
\beqn
\gga(\mz)&=&-\frac{1}{2}\sqrt{\roff} \label{gvega1}\\
\ggv(\mz)&=&\gga(\mz)(1-4\steff) \label{gvega2}
\eeqn
By measuring the ratio $\ggv/\gga$ from various asymmetries,
one determines $\steff$ through eq.(\ref{gvega2})
\beq
\steff=\frac{1}{4}(1-\frac{\ggv}{\gga})
\label{gvsuga}
\eeq
Eqs.(\ref{gvega1}) and (\ref{gvega2}) trade $\ggv$ and $\gga$
for $\steff$ and $\roff$.
In the Standard Model, at tree level, their respective expressions are
\beq
\roff(tree)=1  \;\;\;\;\;\;\;\;\;\; \steff(tree)=\frac{e}{g}
\eeq
($e$ is the electric charge and $g$ the $SU(2)$ weak charge),
but they acquire computable radiative corrections depending, at 1-loop,
linearly on $\mt^2$ and logarithmically on $\mh$ \cite{akv}. In order to
check the Standard Model predictions for these corrections, one should know
the values of $\mt$ and $\mh$. For the moment, we have a lower
limit on them. From top search at Tevatron, one gets \cite{top}
$\mt>113$GeV (95$\%$ of confidence level).
Higgs direct search at LEP
yields a limit $\mh > 63.5$GeV (95$\%$ of confidence level)\cite{hig}.

On the other hand, we can see that LEP limits on possible deviations from
the Standard Model in the values of different observables
give to $\mt$ an upper limit (cf. previous section and section 3).

The forward-backward asymmetry, $\afbf$ is defined through
the angular distribution observed for the fermion $f$ by the
expression
\beq
\frac{d\sigma}{d\cost}=c\{1+\cos^2\theta+\frac{8}{3}\afbf\cost\}
\label{angular}
\eeq
where $c$ is a normalization constant. For $f=\ell$, $\; \theta$ is the
angle between the initial $e^-$ and the final negative lepton $\ell^-$.
$\afbf$ takes into account the angular asymmetry coming from the
parity-violating \Z\ coupling.
At the \Z\ peak, it can be expressed through the $A_f$'s by
\beq
\afbf=\frac{3}{4}A_e A_f
\label{afbf}
\eeq
where $A_e$ and $A_f$ are defined by eq.(\ref{af}). For leptons,
$\ggv/\gga$ is rather small and the lepton forward-backward asymmetry
can be well approximated by
\beq
\afbl=\frac{3}{4}A_e A_{\ell}\simeq 3(\frac{\ggv}{\gga})^2
\label{afbl}
\eeq
Hence, also $\afbl$ is small. The {\it quark} forward-backward
asymmetries are more attractive, since in the Standard Model
$A_b,A_c>A_{\ell}$.
However, hadronization and flavor-identification
problems make the measurement of $\afbb$ and $\afbc$ much more delicate.

Combining eq.(\ref{glep}) and eq.(\ref{afbl}), one can determine
from a measure of $\glep$ (i.e. of $\ggv^2+\gga^2$) and $\afbl$
(i.e. $\ggv/\gga$) the couplings $\ggv$ and $\gga$ separately.
Furthermore, by isolating informations relative to different leptonic
species, one can test the compatibility of results for $e, \mu$ and $\tau$
leptons,
that is lepton universality. Presently, $\gvl$ and $\gal$ universality
is well verified at LEP (within a $1\sigma$ accuracy) \cite{pre}.

At LEP, two methods are used for measuring $\afbf$. One is by fitting
eq.(\ref{angular}),
to the observed experimental distribution. The other is by counting the
number of events in the forward and backward hemisphere
\beq
\afbf=\frac{N_{\sss F}-N_{\sss B}}{N_{\sss F}+N_{\sss B}}
\eeq
The final determination
of $\afb$ is given  after a deconvolution of initial-state QED radiation
effects (in a way analogous to eq.(\ref{radia})) and
other QED effects.
Although the $\afbf$'s have been measured in a wide range of $\rs$ around
the \Z\ peak, in what follows we will concentrate on their values
at $\rs=\mz$.

The four-experiment combined result for $\afbl$ is reported in Table 2
\cite{lef}.
\begin{table}

\[ \begin{array}{|c||c|} \hline
\afbl & 0.0161 \pm 0.0019 \\ \hline
\afbb & 0.098 \pm 0.006 \\ \hline
\afbc & 0.075 \pm 0.015 \\ \hline \hline
A_{\tau} (\ptau)   & 0.138 \pm 0.014 \\ \hline
A_e (\ptaufb)  & 0.130 \pm 0.025 \\ \hline
\alr (SLD) & 0.100 \pm 0.044 \\ \hline
A_{\ell}   & 0.134 \pm 0.012 \\ \hline
\end{array} \]  \label{asy}

\caption{Asymmetries.}

\end{table}
In spite of the smallness of $\afbl$, this quantity has been determined
with an error better than $12\%$.

In principle, a much easier way to measure $A_{\ell}$ is provided
by $\tau$-polarization asymmetries, which depend linearly (and not
quadratically as $\afbl$) on the small parameter $A_{\ell}$.
The $\tau$-polarization asymmetry $\ptau$ is defined as
\beq
\ptau\equiv\frac{\sigma_{\sss R}-\sigma_{\sss L}}{\sigma_{\sss R}
+\sigma_{\sss L}}
\label{ptau}
\eeq
where $\sigma_{\sss R(L)}$ is the cross section for the production of
right(left)-handed $\tau^{\sss-}$.

Averaging over all production angles, $\ptau$ is given by
\beq
\ptau=-A_{\tau}
\label{ptau2}
\eeq
$\ptau$ is measured by fitting momentum distributions
of $\tau$-decay products, which reflect their $\ptau$-dependent
angular distributions in the $\tau$ rest frame.
The four-experiment combined result for $A_{\tau}$ is shown in Table 2.

Studying the angular dependence of $\ptau$ gives further informations
on $A_e$. In fact, one has
\beq
\ptau(\cost)=-\frac{A_{\tau}(1+\cos^2\theta)+2A_e\cost}
{(1+\cos^2\theta)+2A_e A_{\tau}\cost}
\eeq
Therefore, the forward-backward $\ptau$ asymmetry, $\ptaufb$ is
given by
\beq
\ptaufb=-\frac{3}{4}A_e
\label{ptau3}
\eeq
The advantage of a measurement of $\ptaufb$ with respect to $\ptau$
is that  systematic errors from false $\tau$-polarization effects cancel
in the forward-backward subtraction.
The final result for $A_e$ from $\ptaufb$ is reported in Table 2.
Although $\ptaufb$ is presently affected by a rather large error,
its accuracy is expected to improve in the next future.

As is clear from eqs.(\ref{ptau2}) and (\ref{ptau3})
compared to eq.(\ref{afbl}), $\tau$ polarization
has, with respect to $\afbl$, also the advantage of providing a measurement
of the relative sign of $\ggv$ and $\gga$ and of overcoming correlations
between $e, \mu$ and $\tau$ variables. In this way, one can check
lepton universality \cite{pre}.

Another direct and efficient determination of $A_e$ comes from the
measurement of the left-right polarization asymmetry $\alr$ at SLAC by the
SLD experiment \cite{sld} (see Table 2). This is based on the longitudinal
polarization of the initial $e^{\sss-}$ beam.
The accuracy of this result is presently penalized
by the low statistics.

At the end of Table 2, also the average $A_{\ell}$, derived, assuming
lepton universality, by the LEP $A_{\tau}$ and $A_{e}$ measurements
and the SLD $\alr$ determination is reported \cite{lef}.

 As for heavy $b$- and $c$-quark, the measurement of forward-backward
asymmetries gives $\gvb/\gab$ and $\gvc/\gac$.
The main experimental problem here comes from the difficulty of
tagging the heavy quark
in the hadronic background. The LEP outcome for $\afbb$ and $\afbc$ is
shown in Table 2.

In Table 3, a summary of different determinations of $\steff$
comings from various asymmetry measurements is shown \cite{lef}.
\begin{table}

\[ \begin{array}{|c||c|} \hline
       & \steff \\ \hline \hline
\afbl  & 0.2316 \pm 0.0012 \\ \hline
\ptau    & 0.2327 \pm 0.0018 \\ \hline
\ptaufb    & 0.2338 \pm 0.0031 \\ \hline
\afbb  & 0.2322 \pm 0.0011 \\ \hline
\afbc  & 0.2313 \pm 0.0036 \\ \hline
\qfb   & 0.2320 \pm 0.0016 \\ \hline
\alr  & 0.2378 \pm 0.0056 \\ \hline \hline
average  & 0.2322 \pm 0.0006 \\ \hline
\end{array} \]  \label{sineff}

\caption{Different determinations of $\steff$.}

\end{table}
One can note the beautiful agreement among all the determinations
that leads to a final error of only 0.0006 on $\steff$.

In Table 3, also the information coming from charge asymmetry in
\Z\ decays in all hadronic states  is reported.
$\qfb$ is a measurement of the forward-backward asymmetry in the charge
flow in hadronic events. Indeed,  the large value of $\afbf$ for quarks
imply a non-zero average charge produced in the forward and backward
hemispheres. This is given, summing over the five quark flavours, by
\beq
\qfb=\sum_{f} 2 q_f \afbf \frac{\gfer}{\ghad}
\eeq
However, problems connected with hadronic event reconstruction makes
this measurement not as solid as the leptonic ones.

\section{Global fits to the Standard Model}

By considering the bulk of LEP results versus the time,
one finds a constant  progress in accuracy, joined to
a continuous convergence towards the Standard Model predictions.
For instance, previous little discrepancies with the
Standard Model in the \Z\ leptonic width value has faded away in
the last months.

A  true check of consistency for the Standard Model predictions
through the LEP EW data is made complicated by the presence of two
unknown parameters, the top and the Higgs masses, that affect
EW corrections to different observables.
At 1-loop level, EW corrections can be classified in three main
classes \cite{akv}:

$\bullet$ {\it Oblique} or vacuum polarization corrections of vector bosons,
whose main effect is the running of the electromagnetic coupling
constant $\alpha$ from the low scale value ($\alpha=1/137$) up to
\beq
\alpha(\mz)=\left(128.87\pm0.12\right)^{-1}
\label{running}
\eeq
They include also smaller (and calculable) effects due to top and
higgs loops.

$\bullet$ Vertex corrections, that are small and in general uninteresting.
One  exception is the $Zb\bar{b}$ vertex, which is sensitive to
$\mt$.

$\bullet$ Box corrections, that are always very small and negligible.

As a result of the dependence of radiative corrections on
the unknown $\mt$ and $\mh$, one single measurement is not enough
to test the theoretical model, but several and complementary
(i.e. with a different sensibility to $\mt$ and $\mh$) measurements
are necessary.

In practice, what one does is to use as inputs the parameters
that are measured with the best accuracy, that is
$\alpha, G_{\sss F}$ and $\mz$. Then, the Standard Model predictions
(including all the presently available radiative corrections )
for the set of observables that are measured at LEP are computed,
keeping $\mt$ and $\mh$ as free parameters. Finally, the
consistency with the experimental values of these observables is checked,
and ranges of $\mt$ and $\mh$ that give the best fit to them is computed.
By the way, we will see that the present accuracy in LEP data does not yet
allow to constrain $\mh$. Indeed, while EW radiative corrections depend on
$\mt^2$ linearly at the leading order, they are less sensitive to the exact
value of $\mh$, that enters only through $\log \mh$.

The relevant observables for this analysis are $\gz, \sigoh, \rl, \rb$
and $\ggv/\gga$, that is determined from all asymmetries. To these, it is
very convenient to add two other high-precision EW data that are the ratio
of $W$ and $Z$ masses, which is measured at hadron colliders
by the CDF and UA2 groups \cite{cdf}
\beq
\frac{\mw}{\mz}=0.8798\pm0.0028
\label{mwsumz}
\eeq
and the ratio of neutral- and charged-current cross sections in
neutrino-nucleon scattering from CHARM, CDHS and CCFR \cite{cha}
\beq
R_{\nu}\equiv\frac{\sigma^{\sss NC}}{\sigma^{\sss CC}}=0.312\pm0.003
\label{rnu}
\eeq
that gives
\beq
\sin^2 \theta_{W}\equiv 1-\frac{\mw^2}{\mz^2}=0.229\pm0.003\pm0.005
\label{sin}
\eeq
where the first error is experimental and the second
comes from theoretical uncertainties.
Note that the above definition of the Weinberg angle coincides with
$\steff$ (defined by eq.(\ref{gvega2})) only at tree level,
while it  is modified differently by radiative corrections \cite{akv}.

The result of a global fit of all LEP data to theoretical
predictions reveals perfect consistency of the Standard Model
with present data \cite{pre}.
Leaving the strong coupling constant as a free parameter,
and allowing $\mh$ to vary between 60 and 1000 GeV, the values of
$\mt$ and $\alfs$ that give the best fit are reported in the first column
of Table 4, while the effect of including also non-LEP data from
eqs.(\ref{mwsumz}) and (\ref{sin}) in
the analysis are shown in the second column.
\begin{table}

\[ \begin{array}{|c||c|c|} \hline
       & LEP & LEP + COLLIDER + \nu   \\ \hline \hline
\mt (GeV) & 166^{+17+19}_{-19-22} & 164^{+16+18}_{-17-21} \\ \hline
\alfs & 0.120\pm0.006\pm0.002 & 0.123\pm0.006\pm0.002 \\ \hline \hline
\chi^2/d.o.f. &  3.5/8 & 4.4/11 \\ \hline \hline
\steff & 0.2324 \pm0.0005^{+0.0001}_{-0.0002}& 0.2325
      \pm0.0005^{+0.0001}_{-0.0002} \\ \hline
1-\frac{\mw^2}{\mz^2}  & 0.2255 \pm0.0019^{+0.0006}_{-0.0003}
     & 0.2257 \pm0.0017^{+0.0004}_{-0.0003} \\ \hline
\mw (GeV) & 80.25 \pm0.10^{+0.02}_{-0.03}
& 80.24 \pm0.09^{+0.01}_{-0.02}\\ \hline
\end{array} \]  \label{fit}

\caption{Fits to the Standard Model.}
\end{table}
The second error quoted corresponds to the allowed $\mh$ variation.
Also shown in each case
is the (very good) value of the ratio of $\chi^2$ over the number of
degrees of freedom in the analysis.

Although it is not yet possible with present experimental accuracies
to get a significant upper bound on $\mh$ that is more stringent than
the theoretical one of about 1 TeV, a $\chi^2$ analysis prefers low values
for $\mh$ \cite{pre} (even lower than the direct 63.5 GeV limit in a combined
two-variable $\chi^2(\mt,\mh)$ analysis \cite{alt}).

As already stressed in previous sections, the value obtained for $\alfs$
is in perfect agreement with the result $\alfs=0.123\pm0.006$
from hadronic-event-shape analysis.

{}From the same fit,  one also gets the best estimates for other
observables, that are reported in Table 4, as well.
It is very interesting to note that the error obtained
on the $W$ mass (about 100 MeV) is much smaller than
the present experimental error of about 250 MeV coming
from $W$ studies at hadron colliders \cite{alt}.

Even at the present non-conclusive stage, the EW LEP data are very
constraining. In general, one can say, that
the Standard Model has been tested with an accuracy of $0.5\%$ or better.
In the next section, we will discuss the real implications
of this level of accuracy for a decisive test of the theory.

\section{EW data versus  Standard Model ``Born'' predictions}

We want now to discuss  the problem of establishing at which level the
Standard Model radiative-correction pattern is tested by LEP data.
In fact, what one really wants in order to check
the Standard Model is to probe it beyond
what is predicted by the tree-level theory implemented with
QED radiative corrections, the last being the better established part
of the theory. In other words, one would like to detect
{\it purely} EW corrections. Although, the present $0.5\%$ LEP accuracy
seems to be close to this goal, we can see that this aim is not yet
really accomplished.

Pure QED corrections are rather large but computable with good precision.
Their main effects can be accounted for by substituting the
fine structure constant $\alpha$ with its value at the $\mz$ energy scale
$\abar\equiv\alpha(\mz)$ (cf. eq.(\ref{running})) and by computing
initial-state photon radiation effects (cf. section 2.1).

At this point, it is useful to define as ``Born'' predictions \cite{okn}
the Standard Model results obtained by tree-level
calculations, where one uses as input parameters $G_{\sss F}$, $\mz$ and
$\alpha$ is substituted by $\abar$. Accordingly,
the {\it tree-level} Weinberg angle is replaced by
\beq
\sin^2\theta_0\cos^2\theta_0\equiv\frac{\pi \abar}{\sqrt{2}
G_{\sss F} \mz^2}
\eeq
We will call {\it genuine} EW radiation any deviation from these results.

The goal of this procedure gets clear by looking at Table 5 \cite{oku}.
Here, for various observables, the experimental value is reported
in the first column, while the ``Born'' prediction is shown in the
second.
\begin{table}

\[ \begin{array}{|c||c|c|} \hline
Observable & Measured & ``Born"  \\ \hline \hline
\mw/\mz & 0.8798(29) & 0.8768(2) \\ \hline
\gaa & -0.5008(8) & -0.5000 \\ \hline
\gvv/\gaa & 0.0712(28) & 0.0753(12) \\ \hline
\glep \: (MeV) & 83.79(28) & 83.57(2)\\ \hline
\ghad \: (MeV) & 1740(6) & 1741(5)\\ \hline
\gz \: (MeV) & 2489(7) & 2490(5)\\ \hline
\sigoh \: (nb) & 41.55(14)  & 41.44(5) \\ \hline
\rl & 20.77(5) & 20.84(6) \\ \hline
\rb & 0.2191(27) & 0.2197(1) \\ \hline
\end{array} \]  \label{comparison}

\caption{Comparison between measured and ``Born'' values.}
\end{table}
Perfect agreement is found between the two values, in each case,
within errors (reported in brackets).
No deviation from the ``Born'' approximation is observed experimentally.
This means that genuine EW corrections are small and  in particular
compatible with 0, within the present 1$\sigma$ accuracy.
It is important to stress that, on the other hand,
a true Born calculation (which does not include QED effects) deviates
by a few $\sigma$'s by the experimental values.

It has been argued \cite{oku} that the ``smallness'' of EW radiative
corrections
must be due to some {\it conspiracy} between the top quark and other
particles (light quarks, Higgs, $W$ and $Z$ bosons), since
they contribute with different sign to EW loop corrections.
This makes the magnitude of these corrections smaller than their
{\it natural}
value, that is na\"{\i}vely of the order of $\alpha_{\sss W}/\pi\sim\abar$.
Furthermore, it is exactly this smallness that allows to put rather
stringent limits on $\mt$.

Other two years of LEP running are expected to collect a further 50pb\r{-1}
of integrated luminosity and to halve the present statistical errors.
This could be sufficient to start disentangling genuine EW effects.
Furthermore, once the top mass will be measured in a direct way through top
observation at hadron colliders, one should be able to exactly
compute the top contribution to radiative corrections and possibly separate
Higgs loop effects. This could allow to get some more stringent
experimental information on $\mh$, too.

\section{Model independent analysis of LEP data}
The high level of accuracy of present EW data can be also used to
put some constrain on possible new physics.
In order to do that, one has to analyze the experimental data in a model
independent way.
This problem has been by now
extensively studied, following several different approaches \cite{app,bar}.
In general, one parametrizes the possible deviations from the Standard
Model predictions in terms of a set of new variables.
In the approach of ref.(\cite{bar}), one expresses all the basic EW
observables as functions of the four variables
\epuno, \epdue, \eptre\ and \mepb, that vanish in the limit
of the ``Born'' Standard Model approximation.
The fact that, up to now, no observable deviates from the Standard Model
``Born'' predictions implies that presently
these variables are all compatible with 0. Nonetheless, they can be
extracted from data with a precise error, that gives the amount of
experimentally acceptable deviation from Standard Model ``Born''
predictions.
The variables \epuno, \epdue, \eptre\ and \mepb\ include all the
top and Higgs effects. Hence, they
can be extracted from data without referring to particular values
of $\mt$ and $\mh$. In the Standard Model, they can be expressed
as functions of $\mt$ and $\mh$, so that an experimental bound on the
magnitude of the $\varepsilon_i$'s can be translated in top and Higgs mass
limits.

In order to study LEP constraints on new physics corresponding to some
known theoretical model, one can accordingly
 calculate, starting from the same  model, the non-standard contributions
to the $\varepsilon$ parameters. The experimentally acceptable range
for \epuno, \epdue, \eptre\ and \mepb\ can then either exclude or
constrain the new model.

Consequences of the $\varepsilon_i$'s analysis on various {\it fashionable}
extensions of  the Standard Model have been reviewed in ref.(\cite{alt}).

Concerning alternative mechanisms to the EW symmetry breaking,
technicolour models seem to be disfavoured by present data, since they tend
to contribute to \epuno, \eptre\ and \mepb\ more substantially than
experimentally allowed \cite{tec}. On the other hand, this conclusion can be
controverted  by the fact that a true theory of technicolour and realistic
technicolour models have not yet been found. Hence, theoretical predictions
in this case are  rather poorly defined.

A different situation is found for Supersymmetry. Here, we have a well
definite theory which gives clear predictions, although these predictions
depend on several new parameters. In the Minimal
Supersymmetric Standard Model (MSSM), we find two
qualitatively extreme situations \cite{car}.
The first is when all the masses of Susy partners are large.
As far as the $\varepsilon_i$'s analysis is concerned, this case is
equivalent to a Standard Model with a light Higgs ($\mh \ltap 100$GeV)
and, hence, is perfectly compatible with experimental data.
The latter case is when some Susy partners are rather light and close to
their present experimental bounds. For instance, if light gauginos
and s-top quark exist in a range of masses that can be covered by
LEP200 searches, they could produce a detectable deviation from
the Standard Model $\varepsilon$ values.

Finally, the case of extended gauge group has been
considered in the simple case of an extra U(1) \cite{cas}. Present data are
already constraining enough as to allow  only for a very small
amount of mixing ($\xi<1\%$) between the Standard Model \Z\
and the new neutral vector boson
associated to the extra U(1).

\section{Conclusions}

After four years from its starting, LEP has produced a huge amount of
data at the \Z\ peak. These data have allowed the determination of several
EW observables with unprecedented accuracy. There is spectacular
(and improving with time) agreement
with all the pattern of Standard Model predictions,
although the present accuracy  measurements, at the level of $0.5\%$,
 is not yet sufficient to disentangle purely EW radiative
corrections. Nonetheless, the small errors on  different observables
already allow to considerably constrain the top mass in the range
\beq
\mt = \left( 164\pm27 \right) \; GeV
\eeq
Further improvements are foreseen after the '93 and '94 running
completion. With about four times the present statistics, one will
halve the today statistical errors. This hopefully will permit
to distinguish
genuine EW effects and, in case top will be directly observed at Tevatron,
to disentangle Higgs-loop contributions.

On the other hand, by performing a model independent analysis
of the data, one can already put severe constrains on possible
extensions of the Standard Model.

In a pessimistic picture, where one assume that no effect from new physics
will be observed directly in the next future, the comparison between
more and more accurate experimental data and more and more
precise theoretical predictions could be the only way to
discover what is beyond the Standard Model.


\newpage

\begin{thebibliography}{999}

\bibitem{pre}  The LEP Collaborations ALEPH, DELPHI, L3, OPAL and
                The LEP Electroweak Working Group,
                preprint CERN/PPE/93-157(1993).

\bibitem{lef}   J.~Lefran\c{c}ois, plenary talk at the Int. Europhysics
                     Conf. on High Energy Physics, Marseille,
                     July 22-28, 1993.

\bibitem{swa}        talk by M.~Swartz at the
                     XVI Int. Symp. on Lepton-Photon Interactions,
                     Cornell University, Ithaca, N.Y., U.S.A.,
                     August 10-15, 1993.

\bibitem{hol}        W.~Hollik, talk at the
                     XVI Int. Symp. on Lepton-Photon Interactions,
                     Cornell University, Ithaca, N.Y., U.S.A.,
                     August 10-15, 1993.

\bibitem{ber}       F.A.~Berends in ref.\cite{akv}, vol. 1, p. 89.

\bibitem{akv}  \Z\ {\it Phisics at LEP 1},
       G.~Altarelli, R.~Kleiss and C.~Verzegnassi (eds.),
        CERN 89-08, Geneva,1989.

\bibitem{ver}     A.~Blondel, A.~Djouadi and C.~Verzegnassi, preprint
              DESY 92-112 (1992).

\bibitem{top}        talks by P.~Tipton (CDF collaboration) and
                     N.~Hadley (D0 collaboration) at the
                     XVI Int. Symp. on Lepton-Photon Interactions,
                     Cornell University, Ithaca, N.Y., U.S.A.,
                     August 10-15, 1993.

\bibitem{hig}        D.~Treille, plenary talk at the Int. Europhysics
                     Conf. on High Energy Physics, Marseille,
                     July 22-28, 1993.

\bibitem{sld}     The SLD Collaboration, SLAC-PUB-6030(1993);
                  W.~Ash, talk at the Int. Europhysics
                     Conf. on High Energy Physics, Marseille,
                     July 22-28, 1993.

\bibitem{cdf}        CDF Collaboration, F.~Abe \etal, \prl{65}{90}{2243};
                       \prev{D43}{91}2070; \\
                    UA2 Collaboration,\pl{B276}{92}{354}; \\
                    Particle Data Group 1992: \prev{D45}{92}{1}.

\bibitem{cha}   CHARM Collaboration, \pl{B177}{86}{446};
                  \zp{C36}{87}{611};\\
                CDHS Collaboration, \prl{57}{86}{298};\zp{C45}{90}{361}; \\
                 CCFR Collaboration, A.~Bodel, talk at the Int. Europhysics
                     Conf. on High Energy Physics, Marseille,
                     July 22-28, 1993.


\bibitem{alt}        G.~Altarelli, plenary talk at the Int. Europhysics
                     Conf. on High Energy Physics, Marseille,
                     July 22-28, 1993, preprint CERN-TH.7045/93

\bibitem{okn}     V.~Novikov, L.~Okun and M.~ Vysotsky, \np{B397}{93}{35};
                  preprint CERN-TH.6943/93.


\bibitem{oku}        L.B.~Okun, plenary talk at the Int. Europhysics
                     Conf. on High Energy Physics, Marseille,
                     July 22-28, 1993.

\bibitem{app}  M.E.~Peskin and T.~Takeuchi, \prl{65}{90}{964};
                  \prev{D46}{91}{381}; \\
             W.J.~Marciano and J.L.~Rosner, \prl{65}{90}{2963}; \\
            D.C.~Kennedy and P.~Langacker, \prl{65}{90}{2967}.

\bibitem{bar}  G.~Altarelli and R.~Barbieri, \pl{B253}{91}{161};
               G.~Altarelli, R.~Barbieri and S.~Jadach, \np{B369}{92}{3}.

\bibitem{tec}  J.~Ellis, G.L.~Fogli and E.~Lisi, preprint CERN-TH.6383/92
               (1992); \\
           R.S.~Chivukula, S.B.~Selipsky and E.H.~Simmons, \prl{69}{92}{575};
           R.S.~Chivukula \etal, preprint BUHEP-93-11 (1993).

\bibitem{car}    G.~Altarelli, R.~Barbieri and F.~Caravaglios,
                 \pl{B314}{93}{357}.

\bibitem{cas}   G.~Altarelli, R.~Casalbuoni, S.~De~Curtis,
     N.~Di~Bartolomeo, R. Gatto and F.~Ferruglio, preprint
       CERN-TH.5947/93(1993) and reference therein.


\end{thebibliography}
\end{document}